# Novel binding mechanism for ultra-long range molecules

Vera Bendkowsky[1], Björn Butscher[1], Johannes Nipper[1], James P. Shaffer[2], Robert Löw[1], and Tilman Pfau[1]

[1] *5. Physikalisches Institut, Universität Stuttgart, Pfaffenwaldring 57, 70569 Stuttgart, Germany*

[2] *University of Oklahoma, Homer L. Dodge Department of Physics and Astronomy, Norman, OK, 73072 USA*

**Molecular bonds can be divided into four primary types: ionic, covalent, van der Waals and hydrogen bonds. At ultralow temperatures a novel binding type emerges paving the way for novel molecules and ultracold chemistry [1,2]. The underlying mechanism for this new type of chemical bond is low-energy electron scattering of Rydberg electrons from polarisable ground state atoms [3]. This quantum scattering process can generate an attractive potential that is able to bind the ground state atom to the Rydberg atom at a well localized position within the Rydberg electron wave function. The resulting giant molecules can have an internuclear separation of several thousand Bohr radii, which places them among the largest known molecules to date. Their binding energies are much smaller than the Kepler frequencies of the Rydberg electrons i.e. the atomic Rydberg electron state is essentially unchanged by the bound ground state atom. Ultracold and dense samples of atoms enable the creation of these molecules via Rydberg excitation. In this paper we present spectroscopic evidence for the vibrational ground and first excited state of a Rubidium dimer Rb(5S)-Rb(nS). We apply a Born-Oppenheimer model to explain the measured binding energies for principal quantum numbers n between 34 and 40 and extract the s-wave scattering length for electron-Rb(5S)**

**scattering in the relevant low energy regime $E_{kin}$ < 100 meV. We also determine the lifetimes and the polarisabilities of these molecules. P-wave bound states [2], Trimer states [4] as well as bound states for large angular momentum of the Rydberg electron - socalled trilobite molecules [1] - are within reach in the near future and will further refine our conceptual understanding of the chemical bond.**

In a seminal paper published in 1934 Enrico Fermi introduced the notion of scattering length and pseudopotential in the context of low energy electron scattering from polarisable ground state atoms [3]. He realized that despite the fact that the $-\alpha/2|\vec{r}-\vec{R}|^4$ scattering potential[1] between an electron at position $\vec{r}$ and an atom at position $\vec{R}$ is always attractive the quantum mechanical s-wave scattering at low energies can give rise to an either positive or negative scattering length $a$ depending on the relative phase between the ingoing and the scattered electron wave. In this case the interaction is effectively described by the pseudo potential

$$V_{pseudo}(\vec{r},\vec{R}) = 2\pi \cdot a \delta(\vec{r}-\vec{R}), \qquad (1)$$

where $\delta(\vec{r}-\vec{R})$ is the Dirac delta function. Based on Fermi's approach Greene et al. predicted a novel molecular binding mechanism [1]: a low energy Rydberg electron is scattered from an atom with negative scattering length and the resulting attractive interaction binds the Rydberg atom and the ground state atom. As negative s-wave scattering lengths are predicted for alkali atoms [8], alkali dimers are suitable candidates for the search for the resulting ultra-long-range molecules.

The requirements for the validity of Fermi's approach to the interaction between the electron and the ground state atom is that the binding energy is smaller than the Kepler frequency of the Rydberg electron and that the size of the electron wave function $3/2\,n^2$ is much larger than the range of interaction $\tilde{r} = \sqrt{\alpha}$ [7]. Averaged over many scattering events and weighted with the local electron density $|\Psi_{n,l,m}(\vec{R})|^2$ this effectively leads to a mean field potential (mf) between the scattering partners

$$V_{mf}(\vec{R}) = 2\pi \cdot a[k(\vec{R})] \cdot |\Psi_{n,l,m}(\vec{R})|^2 \qquad (2)$$

which again can be repulsive ($a>0$) or attractive ($a<0$). Here, in a semi-classical approximation, the scattering length $a$ is a function of the relative momentum $k(\vec{R})$ of the two scattering partners. The $k$-dependence has been calculated to be $a(k) = a_{atom} + \frac{\pi}{3}\alpha \cdot k + \mathrm{O}(k^2)$, where $a_{atom}$ is the zero-energy scattering length and $\alpha$ the polarisability of the ground state atom [8]. The scattering length, $a$, depends on $\vec{R}$ because the Rydberg electron momentum $k$ changes with its position in the Coulomb potential of the nucleus. Due to the correspondence principle for large principal quantum numbers, $n$, a first approximation for $k(\vec{R})$ is the classical equation given in ref. [1]:

$$\frac{k^2(R)}{2} = -\frac{1}{2n^2} + \frac{1}{R} \qquad (3).$$

---

[1] $\alpha$ is the atomic ground state polarisability. Note that throughout this paper we are using atomic units (a.u.).

In this paper, we restrict ourselves to the simplest Rydberg state namely the S-state ($l=0$). Figure 1 shows the mean field potential given by Eq. (2) together with the electron probability density of the $^{87}$Rb(35S) state as calculated using a Numerov method including quantum defect corrections [9,10]. The included energy levels and wave functions of the molecular potential are computed by a numerical solver [11]. As the molecular potential $V_{mf}(R)$ is proportional to the Rydberg electron probability density, the expected bond length is given by the size of the Rydberg wave function which is 1900 Bohr radii for the $^{87}$Rb(35S) state.

To observe the above mentioned $^3\Sigma$ ultra-long-range molecules we prepare a spin polarised magnetically trapped sample of ultracold $^{87}$Rb atoms in the state $5S_{1/2}$, $m_F=2$ at temperatures T=3.5 µK and peak densities of $n_g=1.5\ 10^{13}$cm$^{-3}$. The sample is trapped in a Ioffe-Pritchard trap with a magnetic offset field of $B_0 = 0.8$ G. We excite ground state atoms by a narrow band two photon excitation to the ($nS_{1/2}$, $m_F=1/2$) state where n is varied between 34 and 40. The bichromatic cw laser system has an effective 2-photon excitation linewidth of ~1 MHz for the two wavelengths of 780nm ($5S_{1/2}$-$5P_{3/2}$) and 480 nm ($5P_{3/2}$-$nS_{1/2}$). Details of the excitation process can be found in [12]. The bichromatic excitation pulses have a duration of 3 µs. The laser power of the red laser is 800 nW in a $1/e^2$ diameter of 1mm and the power of the blue laser is 50 mW in a $1/e^2$ diameter of 80 µm which are both larger than the size of the sample (28 µm). The detuning to the intermediate $5P_{3/2}$ level is larger than 400 MHz to avoid resonant scattering. Polarisations are chosen such that the Rydberg electron spin is conserved [12] with respect to the ground state atoms. For this situation only triplet bound states are expected. Rydberg excited atoms and molecules are detected by field ionization and subsequent ion detection on a multi-channel plate [12]. To correct the binding energies for the Zeeman splitting, the magnetic offset field $B_0$ was determined for each spectrum by an independent measurement.

The spectra obtained for different principal quantum numbers but otherwise essentially comparable conditions are presented in Figure 2. The origin of the spectrum is chosen at the atomic Rydberg line $^{87}$Rb($nS_{1/2}$). On the red side, a slightly broadened peak is observed which is shifted due to the Zeeman effect of the offset field of the magnetic trap $B_0$. Due to the residual inhomogeneous magnetic field direction of the trapping field thermally excited atoms can undergo a spin flip in Rydberg excitation producing this "shoulder". Further to the red side, several lines are observed which we assign to the ultra-long range molecules described above.

Let us first concentrate on the lines at the highest binding energies which we assign to the vibrational ground state of the triplet $^3\Sigma(5S-nS)$ molecule. From our peak densities $n_g$ we expect the Franck Condon factor for the excitation of two free ground state atoms to the bound molecular state to be on the order of $10^{-2}$. But as the excitation of the atomic Rydberg states is highly suppressed by the blockade due to the van der Waals interaction [13], the Franck-Condon factors cannot directly be derived from the relative line intensities.

To assign the observed spectral lines for different principal quantum numbers, we apply the theory based on the described Fermi-Greene model. As the ground state polarisability of Rubidium is accurately known to be $\alpha$=319(6) a.u. [14], the only free parameter in the model is the triplet scattering length $a_{Rb}$. Theoretical predictions for $a_{Rb}$ range between -13 and -17 Bohr radii for the triplet and +0.6 and +2.0 Bohr radii for the singlet case [8]. The results of the calculation are compared with the measured binding energies $E_B$ in Figure 3. The description of the n dependence is surprisingly good given the approximate nature of the model. The best agreement between the binding energies of the $^3\Sigma(5S-nS)(v=0)$ states and the model is obtained for $a_{Rb}$=-18.5 a.u. which is close to the predictions [8]. To our knowledge this is the first measurement of the electron-atom scattering length at such low energies ($E_{kin}$ < 100 meV) which is an important parameter in the context of e.g. electron solvation [15] and structure calculations of negative ions [16]. Note however that our result is dependent upon the model used to calculate the mean field potential and a precision analysis might benefit from an extension of the Fermi-Greene model [17,18]. Based on the good agreement for the vibrational ground state, we apply our calculation to the first vibrational state and assign two observed lines in the 35S and 36S spectra to the ν =1 bound states of those potentials as indicated in Fig. 2 and 3. It has to be considered that the assignment of the ν=1 states can be affected by the inclusion of p-wave scattering of the electron. Also to date the four additional lines remain unassigned and will be the subject of further study. However, calculations predict the existence of similar molecules resulting from p-wave scattering of the Rydberg electron off the ground state atom [2]. This is one possible explanation for the additional lines but has to be confirmed by nontrivial theoretical calculations.

Aside from the vibrational modes, the $^3\Sigma$(5S-nS) molecules also have a rotational structure which is determined only by their mass and bond lengths. The rotational constants range from 11.5 kHz for the $^3\Sigma$(5S-35S) molecule to 9.0 kHz for the $^3\Sigma$(5S-37S) molecule and are thus far below the resolution of the present measurements.

We further characterize the molecular states by lifetime and Stark effect measurements of the molecular ground state ν =0. In Table 1 the measured lifetimes are compared with those of the atomic Rydberg states. For the Rydberg atoms, the lifetimes are slightly longer than the expected free space ones because the metallic environment of the chamber changes the spectral mode density. However, more importantly, the essential result of this measurement is that the molecular lines decay typically a factor of 3-4 faster than the atomic states. The mechanism for this additional molecular decay channel is the subject of further study.

Finally, the investigation of the Stark effect of the atomic 35S state and the molecular $^3\Sigma$(5S-35S)( ν =0) state is shown in Fig. 4. As expected both show a quadratic Stark shift with the electric field. The relative polarisabilities of the atomic and the molecular ν =0 states are $\alpha$ = 1542(7) $10^7$ a.u. and $\alpha$ = 1524(4) $10^7$ a.u. respectively. Absolute values for the polarisability are further affected by a systematic error of 12%. The fact that the values are very similar supports the assumption of the model that the bound ground state atom does not perturb the Rydberg wave function significantly.

To summarize, we have observed the vibrational ground and first excited state of s-wave bound giant Rubidium $^3\Sigma$ (5S-nS) dimers. For the principal quantum numbers reported here (between 34 and 40) their internuclear separation is up to 2556 a.u.. Compared to long range molecules based on resonant dipole-dipole and fine structure interaction first proposed by Stwalley et al. [19, 20] the inner turning point of the vibrational states described in this work is at least one order of magnitude further away from the nucleus. The measured spectra confirm the novel underlying binding mechanisms by the good agreement between theoretically predicted and measured binding energies for a range of principal quantum numbers. As first investigations, we measured the polarisability and the lifetime of these molecules. Future theoretical investigation will clarify the nature of the additional bound states contained in the spectra. Furthermore trimer states [4] as well as high *l* Rydberg states – so called "trilobite-molecules" [1] - will be studied. We speculate that coherent superposition states between free and bound atoms might be possible as the coherent excitation of Rydberg atoms has made significant progress recently [13, 21]. Therefore, the emerging field of ultracold Rydberg chemistry might be enriched by coherent manipulation tools.

We acknowledge financial support from the DFG within the SFB/TRR21 and under the contract PF 381/4-1 and the Landesstiftung Baden-Württemberg, B. B. acknowledges support from the Carl Zeiss foundation. J. P. S. acknowledges support by the Alexander von Humboldt foundation. We thank P. Kollmann for his contribution in the early stage of the experiment.

Correspondence should be addressed to V. Bendkowsky (v.bendkowsky@physik.uni-stuttgart.de) or T. Pfau (t.pfau@physik.uni-stuttgart.de).

Figure 1

Electron probability density in the $(R,\varphi)$-plane $\frac{R}{2\pi}|\Psi_{35,0,0}(R)|^2$ and molecular potential for the $^3\Sigma$(5S-35S) state. The surface plot shows the spherically symmetric density distribution of the Rydberg electron. The molecular potential (green line) is modeled for a polarisability $\alpha$=319 a.u. and a scattering length $a_{Rb}$=-18.5 Bohr radii. Not shown is the repulsive part of the potential for $R$<500 Bohr radii resulting from a zero crossing in the scattering length a(k(R)) at approximately 500 Bohr radii. The attractive part of the potential possesses two bound states in the outermost potential wells at $R$=1900 Bohr radii with binding energies of $E_B(\nu=0)$ = -23.4 MHz and $E_B(\nu=1)$ = -10.6 MHz.

Figure 2

Spectra of the Rydberg states 35S to 37S. The overview spectra on the right side are centered around the atomic Rydberg lines (5S, $m_s=1/2$)→(nS, $m_s=1/2$). The additional "shoulder" at ~ -3 MHz corresponds to the magnetic field dependent transitions (5S, $m_s=1/2$)→(nS, $m_s=-1/2$). On the left side, the observed molecular lines are shown at higher resolution. We assign the left most line of each spectrum to the $^3\Sigma$(5S-nS)($\nu=0$) bound state. Not yet assigned peaks are marked by (♦). All spectra are the averages of up to 30 single spectra and the error bars correspond to 95% confidence bounds on the means.

Figure 3

Measured and calculated binding energies $E_B$ for principal quantum numbers n in the range of 34 to 40. The solid lines are the calculated binding energies for the $\nu=0,1$ states based on the Fermi-Greene model assuming a scattering length $a_{Rb}$ = -18.5 a.u.. The blue shaded area shows the theory for $a_{Rb}$ = -18.5 ± 0.5 a.u.. Red and green symbols represent the measured line centres for the molecular $\nu=0,1$ and the unassigned states respectively. The inset illustrates the definition of the binding energy $E_B$ as the energy difference between the line centers of the atomic doublet $^2$(nS) and the molecular triplet $^3\Sigma$(5S-nS) lines in a finite magnetic field $B_0$ which leads to a Zeeman splitting $\Delta_B$.

Figure 4

Stark map of the atomic (right) and the molecular line for the $^3\Sigma$(5S-nS)($v$ =0) state. The line centres of both the atomic and the molecular state represented by the symbols show a quadratic Stark effect. Their polarisabilities are determined to be $\alpha$ = 1542(7) $10^7$ a.u. and $\alpha$ = 1524(4) $10^7$ a.u. for the atomic 35S state and the $^3\Sigma$(5S-35S)($v$ =0) state respectively (white lines).

**Table 1: Lifetimes $\tau$ of atomic Rydberg state nS and molecular ground state $^3\Sigma$(5S-nS)( $v$ =0)**

| Rydberg state | $\tau_{atom}$ (µs) | $\tau_{molecule}$ (µs) |
|---|---|---|
| 35S | 65 ± 9 | 15 ± 2 |
| 36S | 57 ± 7 | 17 ± 4 |
| 37S | 57 ± 7 | 18 ± 6 |

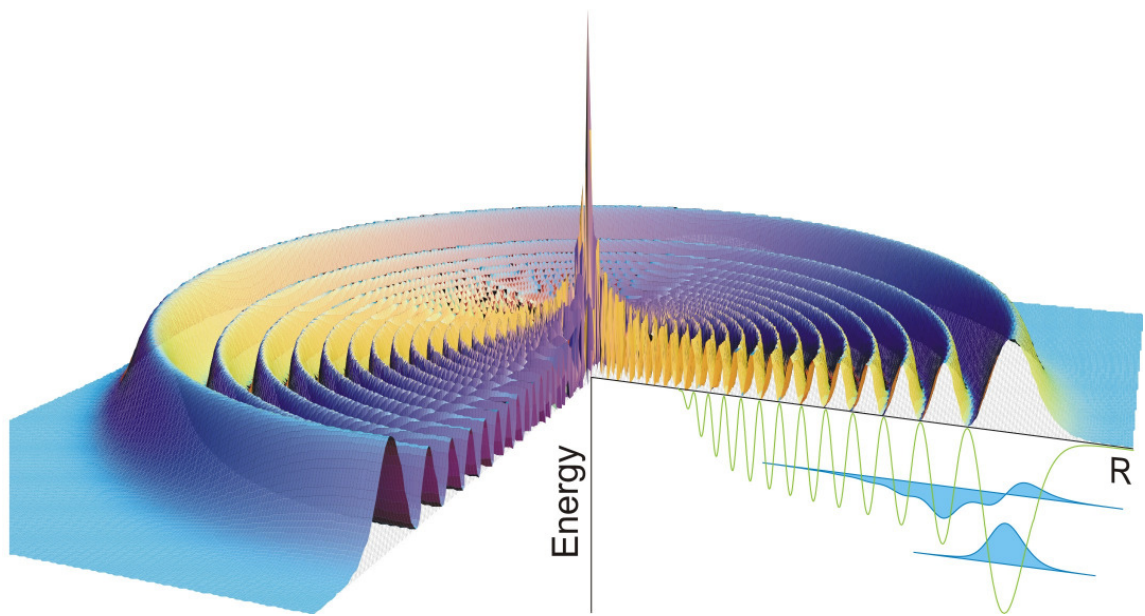

Figure 1

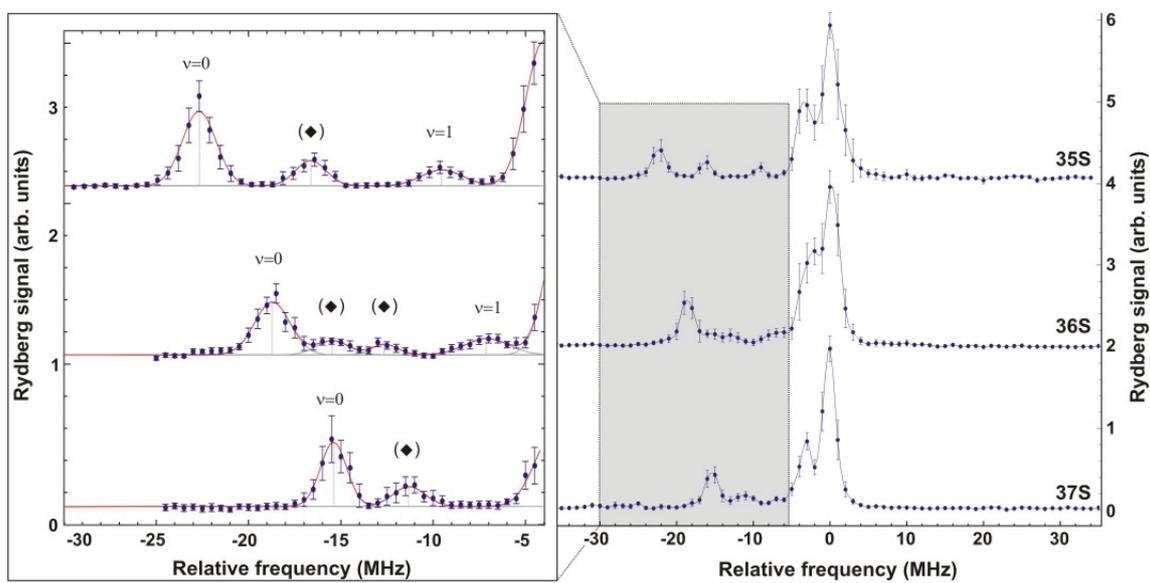

Figure 2

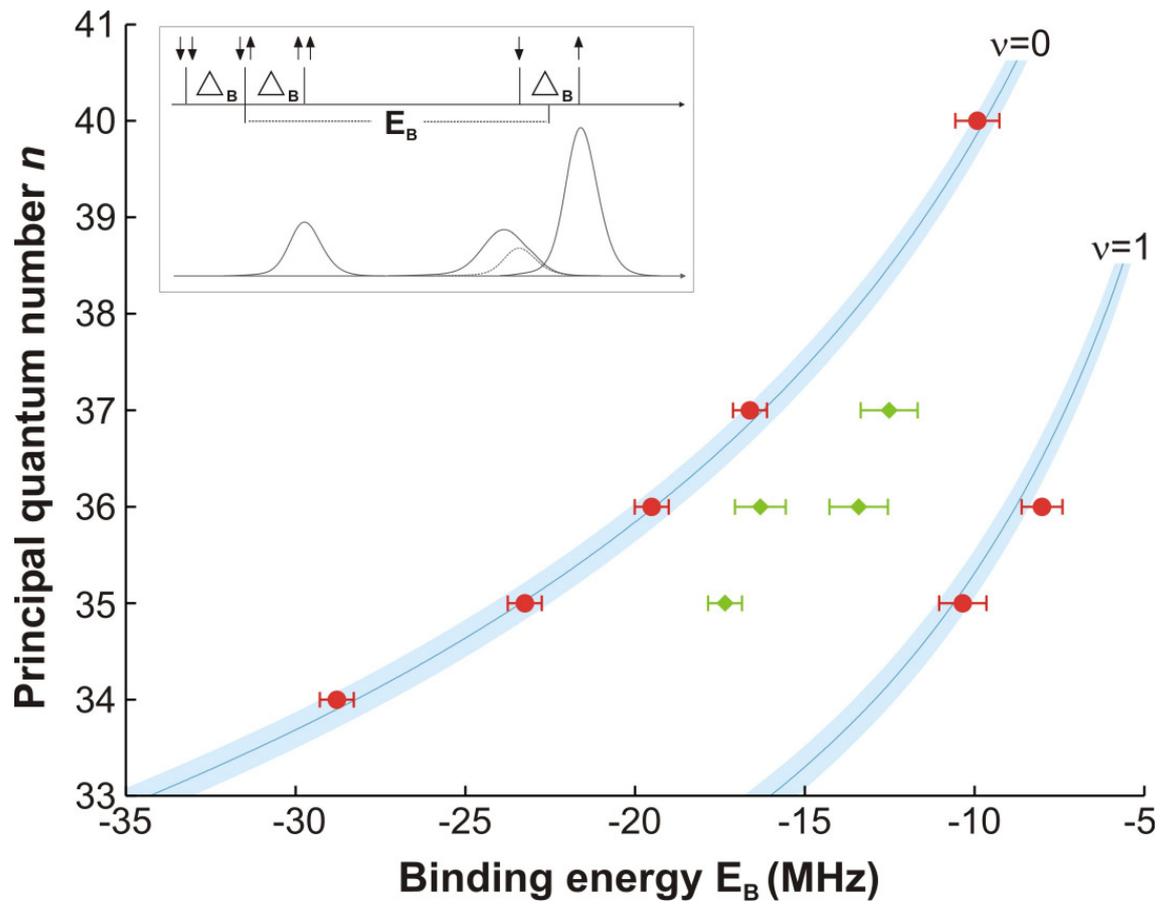

Figure 3

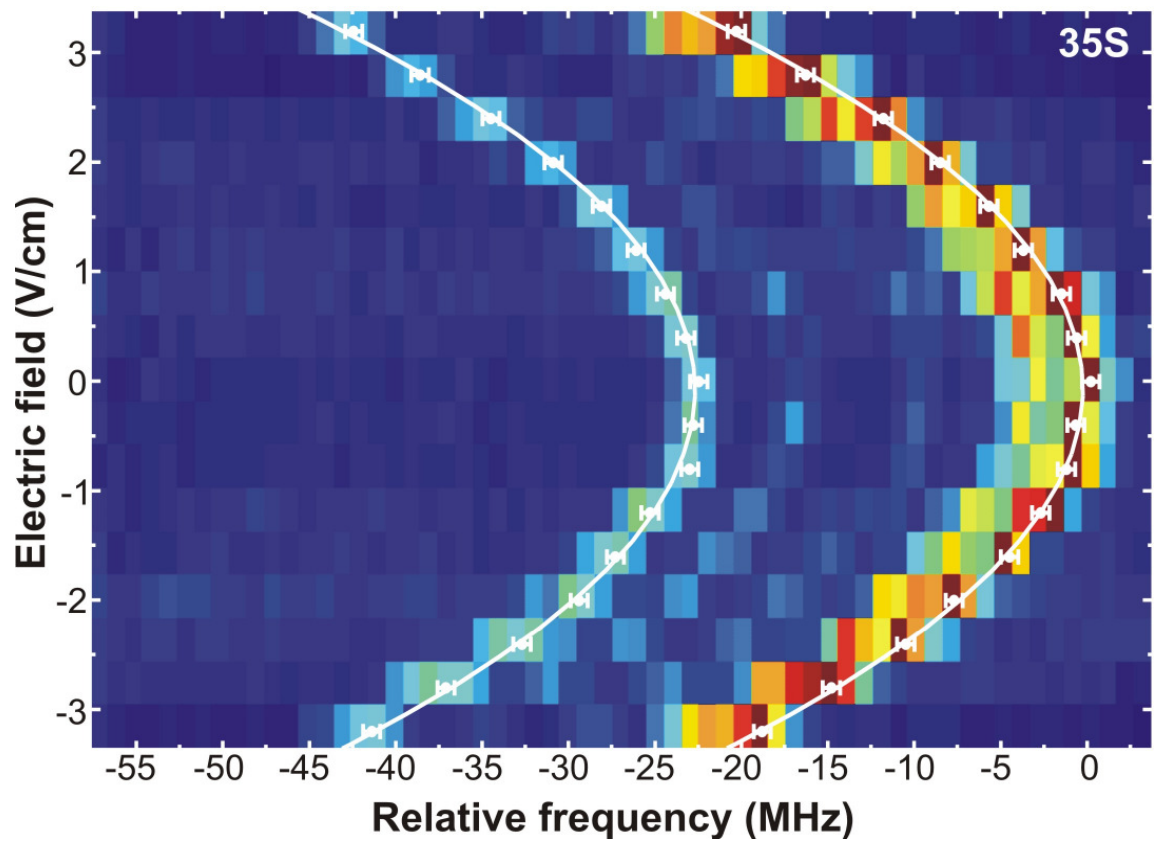

Figure 4